\documentclass[useAMS,usenatbib]{article}
\usepackage{graphicx}

\begin{document}

\noindent
{\bf \large Finding Periods in High Mass X-Ray Binaries}
\\

\noindent{\bf Gordon E. Sarty}\\
{\em University of Saskatchewan, Departments of Psychology and Physics and Engineering Physics,
9 Campus Drive, 
Saskatoon, Saskatchewan S7N 5A5, Canada\\ E-Mail: gordon.sarty@usask.ca}\\ \\
{\bf L\'{a}szl\'{o} L. Kiss}\\
{\bf Helen M. Johnston}\\
{\em School of Physics A28, University of Sydney, New South Wales 2006, Australia}\\ \\
{\bf Richard Huziak}\\
{\em American Association of Variable Star Observers, 25 Birch Street, Cambridge, MA 02138, USA}\\ \\
{\bf Kinwah Wu}\\
{\em Mullard Space Science Laboratory, Holmbury St. Mary, Dorking,
Surrey RH5 6NT, United Kingdom}
\\
\\

\noindent
{\bf Abstract}
This is a call for amateur astronomers who have the equipment and experience for producing high quality
photometry to contribute to a program of finding periods in the optical light curves of high mass X-ray
binaries (HMXB). HMXBs are binary stars in which the lighter star is a neutron star or a black hole
and the more massive star is an O type supergiant or a Be type main sequence star. Matter is transferred
from the ordinary star to the compact object and X-rays are produced as the the gravitational energy of
the accreting gas is converted into light. HMXBs are very bright, many are brighter than 10th magnitude,
and so make perfect targets for experienced amateur astronomers with photometry capable CCD equipment coupled
with almost any size telescope.
\\

\noindent
{\bf 1. Introduction}\\

The brightest class of X-ray sources in the sky are the X-ray binaries (White {\em et al.} 1995).
As such, they were among the first X-ray objects to be studied when sounding rockets and balloon flights
first carried X-ray detectors above the Earth's atmosphere in the 1960s (Begelman and Rees 1998). X-ray binaries consist of a neutron
star or black hole accreting material from a massive companion star. The two main classes of X-ray binaries are
high mass X-ray binaries (HMXBs) and low mass X-ray binaries (LMXBs) with the difference being defined by whether
the companion star is of higher mass or of lower mass than the accreting neutron star or black hole. 

Low mass
X-ray binaries are similar to the cataclysmic variables (CVs) that many AAVSO members observe nightly, except that a white
dwarf, and not a neutron star or black hole, is accreting matter from the companion star in the case of a CV. The companion star
for both CVs and LMXBs are generally dim main sequence M or K stars. CVs tend to be brighter than LMXBs because their accretion
disks are brighter at optical wavelengths. LMXBs are
dim blue objects, optically, with magnitudes typically in the high teens and fainter.

High mass X-ray binaries, on the other hand, tend to be bright ($\gamma$ Cas is one!) because the companion stars are
bright O or B spectral type stars. Early-type O and B stars have lifetimes of only a few million years on the main sequence
so HMXBs are young objects in which one member of the binary has already gone supernova. The current generation of stars (population I) were formed from gas clouds within the disk of the
Milky Way galaxy. Since the HMXBs are young members of the current generation, they have not yet had time to move far from
their birth place in the galactic disk and so are found within the Milky Way on the sky. There are also HMXB sources accessible to amateur astronomers in
the Magellanic Clouds with magnitudes in the 12 to 14 range.

This paper is a call for interested amateur astronomers to participate in a project to find periods in the light curves
of HMXBs. Only 47 of the 130 known HMXBs have known periods (Liu, van Paradijs and van den Heuvel 2000 [hereafter LPH], Bosch-Ramon {\em et al.} 2005). These known periods
correspond to orbital periods, but it is entirely possible that other periods such a star rotation rate could be found. Many of the orbital periods could be a year or two in length making this a long-term project. In
the rest of this paper we will review a little of what is currently known about HMXBs, give a list of objects with unknown
periods, and give some direction about how to become involved in the project.
\\

\noindent
{\bf 2. What are HMXBs?}\\

\begin{figure}
\vspace*{5.5in}


\includegraphics{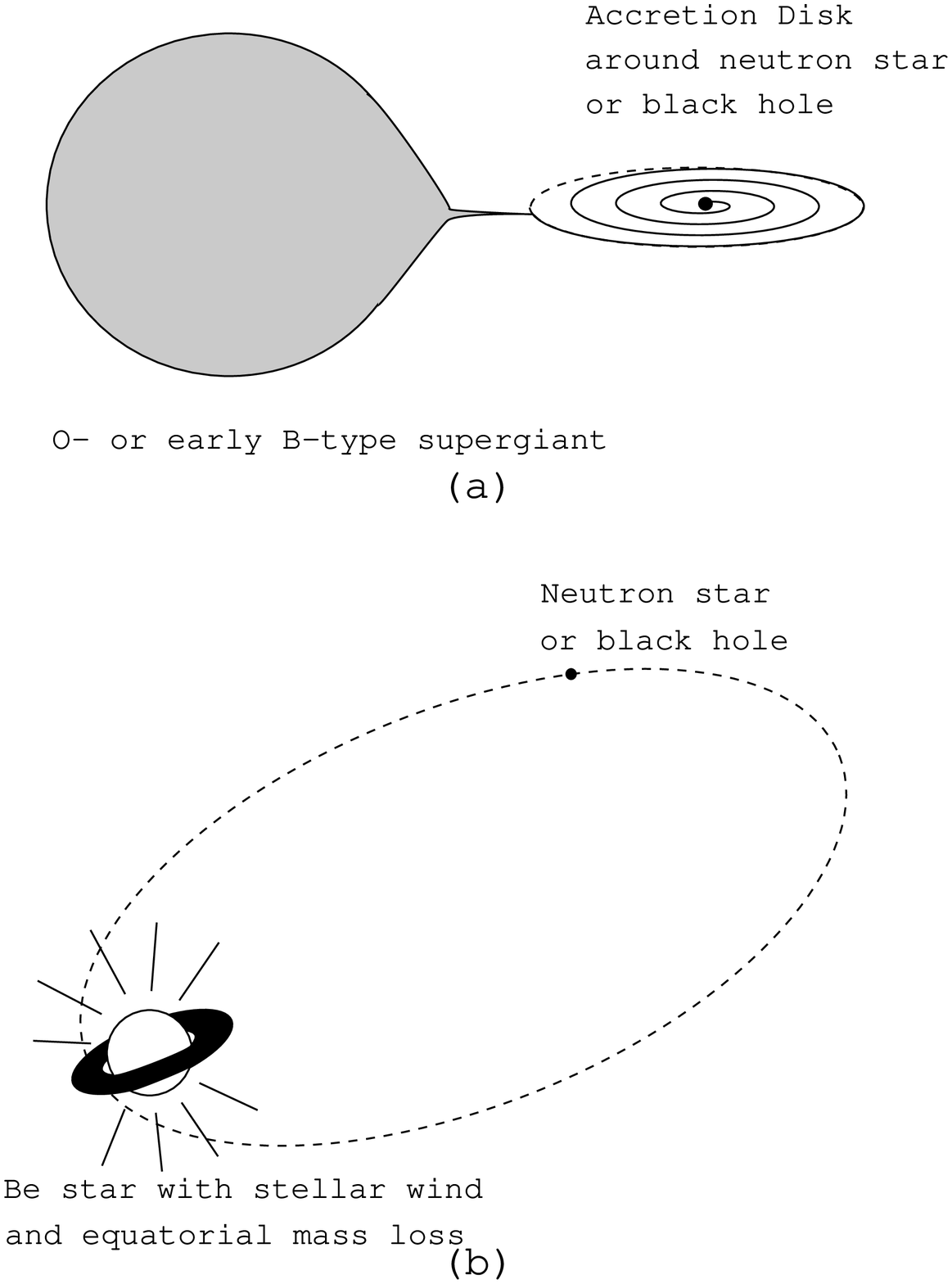}
\caption{Schematics showing the two main types of high mass X-ray binaries. (a) A short period (days)
SG/X-ray binary consisting of a supergiant O- or early-B type star filling or nearly filling
its Roche lobe and transferring mass via an accretion disk to an orbiting neutron star
or black hole. (b) A long period (several weeks to several years) Be/X-ray binary consisting of
a Be rapidly rotating main sequence star that transfers mass to a neutron star
or black hole in a highly eccentric orbit via mass loss through stellar wind and
through a centrifugally produced equatorial disk.  \label{fig1}}
\end{figure}

There are two broad classes of HMXBs (see Figure \ref{fig1}). Those are the short period ones (a few days) with O- or early B-type supergiant companions (SG/X-ray binary) and longer period ones
(from several weeks to several years) with B type companions (Be/X-ray binary) (Verbunt and van den Heuvel 1995).
The majority of known systems are Be/X-ray binaries. Both types of HMXBs are thought to have had one
or more episodes of mass transfer between the two original stars before one of them exploded as a type Ib supernova (helium star core
collapse, all the hydrogen would have been transferred to the companion). 

In the case of a short period SG/X-ray binary, also known as a ``standard'' HMXB, the supernova progenitor is theorized to have swollen in diameter to completely engulf the other star and a spiral-in
phase, similar to that hypothesized for CVs, brought the core of the imminent supernova into a short period orbit with its
companion. After the supernova, the short period SG/X-ray binaries, as we now observe them, transfer mass from the now evolved O type supergiant companion
to the neutron star or black hole supernova remnant. The O supergiant will have a mass greater than 15$M_{\odot}$. The companion star will fill or nearly fill its Roche lobe and the mass
transfer rates will be high enough to produce a permanently bright X-ray binary. The famous Cygnus X-1 is a
SG/X-ray binary that fills its Roche lobe with an orbital period of 5.6 days. Continued evolution (which is
relatively fast because of the high mass transfer rate in a shrinking orbit) and expansion of the companion
will lead to a second spiral-in phase with the roles of the two stars now reversed. The companion will go supernova in its turn leaving a binary neutron star or a neutron star and a black hole. Through the emission of gravitational radiation, that exotic
binary will eventually merge to produce a short duration gamma-ray burst (Gehrels {\em et al.} 2005). Some short period HMXBs exhibit relativistic jet outflow
as a result of the mass transfer. When jets are present, the HMXB is commonly referred to as a microquasar because it is,
in many ways, a
miniature version of a quasar at the heart of some active galactic nuclei (AGN) (Wu {\em et al.} 2002).

In the case of a long period Be/X-ray binary, the original orbit was wide enough to prevent a spiral-in phase before the first supernova.
In that case no mass or angular momentum was lost from the system and the mass transfer eventually caused the orbit to
widen even further. The sideways kick of the supernova also caused the current orbit to be highly eccentric. The companion
star will generally be a rapidly rotating Be (e for `emission' spectrum) star. The massive B star, with a mass between
8 and 15$M_{\odot}$, loses much mass through its
stellar wind and also through an equatorial disk caused by the star's rapid rotation. The equatorial disk is the
source of the emission lines in the star's spectrum. Gas lost through both mechanisms will
accrete onto the neutron star or black hole to produce X-rays. Occasional increases in mass flow from the equatorial bulge can lead to very
bright X-ray transients. Also mass transfer and the system's brightness will increase as the neutron star or black hole
passes periastron. In fact, many Be/X-ray binaries are generally visible in the X-ray, as X-ray transients, only during
periastron passage. The originally closer Be/X-ray binaries will evolve to a merger after the Be star moves off the main
sequence, fills its Roche lobe, and spirals in toward the compact star. Originally wider Be/X-ray binaries will evolve into binary radio pulsars (Verbunt and van der Heuvel 1995)
like the famous gravitational wave emitting Hulse-Taylor pulsar. The gravitational radiation will cause the radio pulsar
to eventually also merge in a short duration gamma-ray burst.

In searching for periods in HMXBs we primarily expect to find orbital periods. But finding other periods, such as the spin
rate of the compact star, is also a possibility, especially if the compact star turns out to be a white dwarf (WD). A system containing
a white dwarf would not be an SG or Be X-ray binary but would be a completely different kind of object, one
intermediate between a CV and the standard HMXB. X-ray systems with white dwarfs are thought to be in a permanent
nova situation, with continuous nuclear fusion on the surface of the WD of the accreted gas, and are thought to be associated
with objects having an extreme ultra-soft (EUS) X-ray spectrum (also known as ``super-soft'' X-ray sources (Hellier 2001)). 
HMXB-like systems that turn out to have white dwarfs instead of neutron stars or black holes also exist; an example is $\gamma$ Cas.

Orbital motion can cause periodic variation in a HMXB light curve in a number of ways. For a short period SG/X-ray binary
that fills or nearly fills its Roche lobe, there will be ``ellipsoidal variation'' caused by the changing projected area of the
star as seen from Earth. For the short period binaries, and even for longer period binaries with a sufficiently
massive black hole (Copperwheat {\em et al.} 2005), irradiative heating of the companion star by X-rays from the accretion disk around the compact star can cause light curve variation as the brighter heated side of the star becomes visible and hidden
in the course of the orbit. Finally, the increased accretion rate caused by the passage of the compact star through
periastron around the Be star can cause a brightening in the light curve.

HMXBs tend to be relatively active objects and other effects can give rise to variability in the optical light curve. Optically,
most of the light comes from the O or B star. In the X-ray region of the electromagnetic spectrum the light comes
from the accretion disk and accretion layer on the surface of neutron star. The X-ray spectrum from the accretion layer
is hard (bright at high energies) while the X-ray spectrum from the inner disk is soft (bright at lower energies). So the
absence of a hard X-ray spectra can be evidence for a black hole because a black hole has no surface and hence
no accretion layer. In any case, variation of accretion rate will cause
variation in X-ray brightness which, in turn, will cause variation in optical light produced by the ``reprocessing'' of
the X-ray radiation in either the accretion disk or in the atmosphere of the companion star. Variations in the accretion rate
can be quite dramatic if the accretion rate is large enough to produce radiation at the ``Eddington limit''. If the
accretion is at the Eddington limit, pressure from the emitted light radiation will be high enough to push the
accreting gas away and shut off the mass flow. With the mass flow cut off, the disk and accretion layer luminosity
then falls below the Eddington limit and mass flow starts again. If the system possesses a jet then more complicated behaviour
is possible and wide variability in the X-ray light curve is common (Muno {\em et al.} 1999), which may or may not
translate into optical variability. These other sources of variability may make it difficult to extract the period
from some of the HMXB light curves.
\\

\noindent
{\bf 3. List of targets}\\

Table \ref{table1} gives a list of our HMXB program stars. They are the objects listed by
LPH that have unknown orbital periods and known optical counterparts. There are 47 program stars.
9 are listed as O type, which are likely to have short orbital periods (days), 36 are listed as B type, which
are likely to have long periods (up to 1 or 2 years), and 2 are of unspecified type.

The columns in Table \ref{table1} give information as follows. The first column gives the entry number in order
of appearance in the LPH catalogue. The second column gives the name of the object as given in the LPH catalogue.
Most objects have several names and will be given Harvard designations by the AAVSO as data are posted to the
AAVSO international database. The third column gives the V magnitude of the optical counterpart (the O or B star) of the
X-ray binary. The fourth column gives the spectral type, the fifth column gives the type, P, T or U which mean:
\begin{itemize}
\item P: X-ray pulsar. If the object is an X-ray pulsar, the pulse period is given in the last column of
Table \ref{table1}. X-ray pulsars are the X-ray binary analogue of CV polars. They have high magnetic
fields that channel the flow onto one or two hot spots on the surface of a neutron star. As can be seen, the pulse periods generally imply a neutron star spin rate that is
far too rapid to be detectable by photometric exposures that last tens of seconds. But the possibility of
detecting the longer period spin rates by photometry remains.
\item T: transient X-ray source. These are mostly B type objects, as expected, but 4 of the
transient sources have O companions; however, of those 4, 3 are emission objects suggesting equatorial
mass loss.
\item U: ultra-soft X-ray spectrum. These sources are black hole candidates or if the
X-ray spectrum is ``extreme ultra-soft'' (EUS) the accreting object may be a white dwarf
in which the accretion rate is high enough to support nuclear fusion -- a permanent nova.
\end{itemize}
Here are some notes on the individual objects as given in LPH:\\

\noindent
{\em LPH001}: Small Magellanic Cloud (SMC) object.

\noindent
{\em LPH004}: SMC object as well, SMC X-3, $v_{\mbox{rot}} \sin i \sim$200 km s$^{-1}$.

\noindent
{\em LPH008}: SMC 25.

\noindent
{\em LPH009}: In SMC.

\noindent
{\em LPH010}: SMC 32.

 \noindent
{\em LPH011}: SMC X-2, $v_{\mbox{rot}} \sin i \sim$200 km s$^{-1}$.

\noindent
{\em LPH012}: $\gamma$ Cas, variable Be star, possibly WD sys., $v_{\mbox{rot}} \sin i \sim$300--500 km s$^{-1}$.

\noindent
{\em LPH017}: In SMC.

\noindent
{\em LPH018}: In SMC, EUS object.

\noindent
{\em LPH020}: In SMC, consistent with supernova remnant SNR 0101-724.

\noindent
{\em LPH021}: In SMC.

\noindent
{\em LPH023}: In SMC, consistent with supernova remnant SNR 0104-72.3.

\noindent
{\em LPH024}: In SMC, displays optical outbursts.

\noindent
{\em LPH028}: In SMC, black hole candidate, rotational velocity 145 km s$^{-1}$.

\noindent
{\em LPH029}: In open cluster NGC 663, $v_{\mbox{rot}} \sin i \sim$250 km s$^{-1}$.

\noindent
{\em LPH033}: CI Cam, possible BH candidate, possible symbiotic-type X-ray binary.

\noindent
{\em LPH035}: In Large Magellanic Cloud (LMC). 

\noindent
{\em LPH036}: In LMC.

\noindent
{\em LPH038}: In LMC.

\noindent
{\em LPH039}: In LMC.

\noindent
{\em LPH046}: In LMC, black hole candidate.

\noindent
{\em LPH053}: In LMC.

\noindent
{\em LPH054}: In LMC.

\noindent
{\em LPH055}: In LMC, optical id. not completely certain.

\noindent
{\em LPH056}: In LMC.

\noindent
{\em LPH058}: Possibly the GeV $\gamma$-ray source 2EG J0635+0521. 

\noindent
{\em LPH062}: In open cluster NGC 2516.

\noindent
{\em LPH067}: Possible Wolf-Rayet star + O star rather than a HMXB.

\noindent
{\em LPH069}: Probably the same as LPH068.

\noindent
{\em LPH071}: $v_{\mbox{rot}} \sin i \sim$300 km s$^{-1}$.

\noindent
{\em LPH079}: Possible white dwarf accretor.

\noindent
{\em LPH080}: Possible white dwarf accretor.

\noindent
{\em LPH100}: A $\gamma$-ray emitting persistent microquasar.

\noindent
{\em LPH129}: Herbig Ae/Be candidate. \\

The high rotational velocities of LPH004, 011, 012, 028, 029 and 071 fit with their B-emission type character, as B-emission stars always are very rapid rotators. Such systems are expected to have long orbital periods: several weeks to several years.
The period of LS 5039/RXJ1826.2-1450 (LPH100) has been found to be 3.9060$\pm$0.0002
days (Bosch-Ramon et al. 2005) but we are keeping it on our list because its
light curve to date has shown no significant variation; its period was found spectroscopically.
At high enough precision, there may be some structure to the light curve at optical wavelengths
and variations of $\sim$0.4 mag have been reported in the H and K infrared bands (Clark et
al. 2001).
\\

\noindent
{\bf 4. Observing methods}\\

The amplitudes of the light curves may be very small so it may be necessary to observe with a precision to
0.005 mag or better. However at this point we simply do not know what the amplitude of the variations
will be, so more standard data with 1\% or so errors (0.01 mag) will also be very useful. The finer requirement is similar to the precision needed to detect the transits of extrasolar
planets so the same equipment and observing techniques required for observing extrasolar planet transits
can be used to observe HMXBs. An excellent overview of the equipment and observing techniques required
to achieve 0.005 mag precision is given by Castellano {\em et al.} (2004). The HMXBs generally will need to be observed
through the standard Johnston B, V, R and I filters so that we can easily combine data from different
observers and so we can see if any variable color effects are present. But, again, V-band only observations,
for example, will also be very useful.

We will be organizing the observing efforts into campaigns using the AAVSO-photometry e-mail list. The idea
is to get nearly continuous coverage of the HMXB's light curve with what might be called an
``Amateur Astronomers' Whole-Earth Telescope'' or AAWET. Observers at all longitudes would provide
continuous coverage for a month or so similar to how nearly continuous coverage was obtained in a recent
AAVSO project on SS Cygni. Also in a manner similar to how the SS Cyg campaign was run, observers would
reduce their data and post to the AAVSO's international database in the usual manner. We would also request that observers reduce a few check stars in their frames because the light curves of those check stars
will be necessary to verify any unusual activity seen in the HMXB's light curve. The check star data won't be submitted to the AAVSO database but should be archived by the observer in case the data are needed. Charts showing the field
and preferred check stars will be made available to the observers when a campaign starts.

Once a long enough light curve is produced, it will be subject to a period analysis using standard software such as
T. Vanmunster's {\em Peranso} software ({\tt http://users.skynet.be/fa079980/peranso/index.htm}). Continuous coverage at a reasonably high time resolution will also allow us to search for short periods and to eliminate high frequency noise in the periodogram.

To date we have photometric data for LPH046, 053, 067, 069, 071, 088, 095, 100, 107, 115, 123, 127, 128
and 129. So stars in that list will be our first campaign targets. These are the brighter stars in the list (brighter
than magnitude $\sim$14). Outside of the list of the stars already observed are the fainter ones, which will
be tougher to observe with high precision, and stars that refuse to show simple light curves, like
$\gamma$ Cas.
If you are interested in participating in this challenging project of finding periods for high mass X-ray binaries, keep a look out for observing requests on the AAVSO-photometry e-mail list.
\\

\newpage
\noindent
{\bf References}\\

\noindent
Begelman, M., and Rees, M. 1998, {\em Gravity's Fatal Attraction: Black Holes in the Universe}, Scientific
American Library, New York.\\

\noindent
Bosch-Ramon, V., Paredes, J.M., Rib\'{o}, M., Miller, J.M., Reig, P., and Marti, J. 2005, {\em Astrophys. J.}, {\bf 628}, 388.\\

\noindent
Castellano, T.P., Laughlin, G., Terry, R.S., Kaufman, M., Hubbert, S., Schelbert, G.M., Bohler, D.,
and Rhodes, R. 2004, {\em J. Amer. Assoc. Var. Star Obs.}, {\bf 33}, 1.\\

\noindent
Clark, J.S., Reig, P., Goodwin, S.P., Larionov, V.M., Blay, P., Coe, M.J., Fabregat, J., Negueruela, I.,
Papadakis, I., and Steele, I.A. 2001, {\em Astron. Astrophys.}, {\bf 376}, 476.\\

\noindent
Copperwheat, C., Cropper, M., Soria, R., and Wu, K. 2005, {\em Mon. Not. R. Astron. Soc.}, {\bf 362}, 79.\\

\noindent
Gehrels, N. {\em et al.} 2005, {\em Nature}, {\bf 437}, 851.\\

\noindent
Hellier, C. 2001, {\em Cataclysmic Variable Stars: How and Why They Vary}, Springer, Praxis, Chichester, UK.\\

\noindent
Liu, Q.Z., van Paradijs, J., and van den Heuvel, E.P.J. 2000, {\em Astron. Astrophys. Suppl. Ser.}, {\bf 147}, 25. [LPH]\\

\noindent
Muno, M.P., Morgan, E.H., and Remillard, R.A. 1999, {\em Astrophys. J.}, {\bf 527}, 321.\\

\noindent
Stetson, P.B. 1987, {\em Publ. Astron. Soc. Pacific}, {\bf 99}, 191.\\

\noindent
Verbunt, F., and van den Heuvel, E.P.J. 1995, ``Formation and evolution of neutron stars and black holes in binaries''
in {\em X-Ray Binaries},
eds. W.H.G. Lewin, J. van Paradijs, and E.P.J van den Heuvel, p. 457, Cambridge University Press.\\

\noindent
White, N.E., Nagase,F., and Parmar, A.N. 1995, ``The properties of X-ray binaries'' in {\em X-Ray Binaries},
eds. W.H.G. Lewin, J. van Paradijs, and E.P.J van den Heuvel, p. 1, Cambridge University Press.\\

\noindent
Wu, K., Stevens, J.A., and Hannikainen, D.C. 2002, {\em Publ. Astron. Soc. Aust.}, {\bf 19}, 91.

\begin{center}
\begin{table}
\caption{Program star list. See main text for notes. \label{table1}}
\begin{tabular}{cccccc}
\hline
{\em LPH}	&	{\em Name}	&	{\em V}	&{\em Spectral}	&	{\em Type}	&	{\em Ppulse} \\
{\em No.}   &               &           &   {\em Type}  &               &   {\em (sec)} \\ 
\hline
001	&	J0032.9-7348	&	15.3	&	Be	&		&		\\
004	&	0050-727	&	14.   :	&	O9 III-Ve	&	T	&		\\
008	&	J0051.9-7311	&	14.4	&	Be	&	T	&		\\
009	&	J0052.1-7319	&	14.667	&	Be	&	TP	&	15.3	\\
010	&	J0052.9-7158	&	15.46	&	Be	&	TU	&		\\
011	&	0053-739	&	16	&	B1.5Ve	&	T	&		\\
012	&	0053+604	&	1.6  - 3.0	&	B0.5Ve	&		&		\\
017	&	J0058.2-7231	&	15	&	Be	&		&		\\
018	&	J0059.2-7138	&	14.1	&	B1IIIe	&	TUP	&	2.7632	\\
020	&	J0103-722	&	14.8	&	O9-B1(III-V)	&	P	&	345.2	\\
021	&	0103-762	&	17	&	Be	&	T	&		\\
023	&	J0106.2-7205	&	16.7	&	B2-B5 III-Ve	&		&		\\
024	&	J0111.2-7317	&	15.32	&	B0-2III-Ve	&	TP	&	31.0294	\\
028	&	J0117.6-7330	&	14.19	&	B0.5IIIe	&	TP	&	22.07	\\
029	&	J0146.9+6121	&	11.33	&	B1Ve	&		&	1404.2	\\
033	&	J0421+560	&	9.25	&	--	&	T	&		\\
034	&	J0440.9+4431	&	10.78	&	B0V-IIIe	&	P	&	202.5	\\
035	&	J0501.6-7034	&	14.5	&	B0e	&		&		\\
036	&	J0502.9-6626	&	14.22	&	B0e	&	TP	&	4.0635	\\
038	&	J0516.0-6916	&	15	&	B1e	&	T	&		\\
039	&	J0520.5-6932	&	14.4	&	O8e	&	T	&		\\
040	&	0521+373	&	7.51	&	B0IVpe	&		&		\\
046	&	J0532.5-6551	&	12.3	&	OB	&	U	&		\\
053	&	J0541.4-6936	&	12.01	&	B2 SG	&		&		\\
054	&	J0541.5-6833	&	14.02	&	OB0	&		&		\\
055	&	0544-665	&	15.4	&	B1Ve	&		&		\\
056	&	J0544.1-710	&	15.33	&	Be	&	TP	&	96.08	\\
057	&	0556+286	&	9.2	&	B5ne	&		&		\\
058	&	J0635+0533	&	12.83	&	B2V-B1IIIe	&	P	&	0.0338	\\
061	&	0739-529	&	7.62	&	B7IV-Ve	&		&		\\
062	&	0749-600	&	6.73	&	B8IIIe	&		&		\\
067	&	1024.0-5732	&	12.7	&	O5:	&	P	&	0.061	\\
069	&	J1037.5-5647	&	11.3	&	B0V-IIIe	&	P	&	862	\\
071	&	1118-615	&	12.1	&	O9.5 III-Ve	&	PT	&	405	\\
079	&	1249-637	&	5.31	&	B0IIIe	&		&		\\
080	&	1253-761	&	6.49	&	B7Vne	&		&		\\
081	&	1255-567	&	5.17	&	B5Ve	&		&		\\
088	&	1555-552	&	8.6	&	B2nne	&		&		\\
095	&	J1744.7-2713	&	8.4	&	B2V-IIIe	&		&		\\
\hline
\end{tabular}
\end{table}
\end{center}
\clearpage

\begin{center} 
Table1: continued. 
\begin{tabular}{cccccc}
\hline
{\em LPH}	&	{\em Name}	&	{\em V}	&{\em Spectral}	&	{\em Type}	&	{\em Ppulse} \\
{\em No.}   &               &           &   {\em Type}  &               &   {\em (sec)} \\ 
\hline
100	&	J1826.2-1450	&	11.23	&	O7V((f))	&		&		\\
107	&	1845.0-0433	&	13.96	&	O9.5I	&	T	&		\\
115	&	1936+541	&	9.8	&	Be	&		&		\\
117	&	1947+300	&	14.2	&	--	&	T	&		\\
123	&	J2030.5+4751	&	9.27	&	B0.5V-IIIe	&		&		\\
127	&	2202+501	&	8.8	&	Be	&		&		\\
128	&	2206+543	&	9.9	&	B1e	&		&	392.     ?	\\
129	&	2214+589	&	11	&	Be	&		&		\\
\hline
\end{tabular}
\end{center}

\end{document}